\documentstyle[preprint,aps]{revtex}
\newcommand{\be}{\begin{equation}}
\newcommand{\ee}{\end{equation}}
\newcommand{\ben}{\begin{eqnarray}}
\newcommand{\een}{\end{eqnarray}}

\newcommand{\iii}{\'{\i}}

\begin{document}

\title{Entanglement Distribution and Entangling Power of Quantum Gates}
\author{J. Batle$^1$\footnote{E-mail: 
vdfsjbv4@uib.es, corresponding author.}, M. Casas$^1$, A. Plastino$^2$,  
and A. R. Plastino$^{1,\,2,\,3}$  }
\address{ $^1$Departament de F\iii sica, Universitat de les Illes Balears,
07122 Palma de Mallorca, Spain \\  $^2$National University La Plata and
CONICET,  C.C. 727, 1900 La Plata, Argentina \\ $^3$Department of Physics,
University of Pretoria, Pretoria 0002, South Africa}

\draft

\maketitle

\begin{abstract}

Quantum gates, that  play a fundamental role 
in quantum computation and other quantum 
information processes, are unitary evolution 
operators $\hat U$ that act on a composite 
system changing its entanglement. In the present 
contribution we study some aspects of these 
entanglement changes. By recourse of a Monte Carlo 
procedure, we compute the so called ``entangling power" 
for several paradigmatic quantum gates and 
discuss results concerning the action of the CNOT gate. 
We pay special attention to the distribution of 
entanglement among the several parties involved.

\vskip 5mm

 Pacs: 03.67.-a; 89.70.+c; 03.65.Bz

\vskip 5mm

\noindent  Keywords: Quantum Entanglement; Unitary Transformations; 
Quantum Information Theory

\end{abstract}
\maketitle

\newpage

\section{Introduction}

Entanglement is one of the most fundamental 
aspects of quantum physics \cite{Schro} and can be regarded as a 
physical resource associated with peculiar non-classical 
correlations between separated quantum systems. Entanglement 
lies at the basis of important quantum information processes 
\cite{P93,WC97,BEZ00,Nature2004} such as quantum cryptographic 
key distribution, quantum teleportation, superdense coding and 
quantum computation \cite{EJ96,galindo,CiracPhysTod}.

Quantum gates, the quantum generalization of the so-called standard 
logical gates, play a fundamental role in quantum 
computation and other quantum information processes, being 
described by unitary transformations $\hat U$ acting on 
the relevant Hilbert space (usually, that for a multi-qubit system). 
In general, a quantum gate acting on a composite system changes 
the entanglement of the system's concomitant quantum state. It 
is then a matter of interest to obtain a detailed characterization of the 
aforementioned entanglement changes (\cite{Z01,KC01,DVCLP01,BPCP03}). 
To such an end, the study is greatly simplified if one is able to 
conveniently parameterize the gate's non-local features. We need 
$N^2-1$ parameters to describe a unitary transformation $U(N)$ in a system 
of $N = N_A \times N_B$ dimensions. In the case of two qubits ($N=2\times2$) 
one needs just three parameters ${\bf \lambda} 
\equiv (\lambda_1,\lambda_2,\lambda_3)$ (\cite{KC01}), since any 
two-qubits quantum gate can be decomposed in the form of a product of local 
unitarities, acting on both parties, and a ``nuclear" part $\tilde U$, 
which is completely non-local. Given a quantum gate $U$, the 
concomitant distribution of entanglement changes is equivalent, on average, 
to the one produced by $\tilde U$, and we need to know the vector 
${\bf \lambda}$.

In addition to studying  changes in the entanglement of a given state produced 
by quantum gates, we would like  to ascertain  entangling 
capabilities of  unitary operations or evolutions. In point of fact, 
the latter enterprise complements the former. By looking at the distribution 
of entanglement changes induced by several quantum gates, 
one can deduce a special formula that quantifies the 
``entangling power". To such an end we use the definition introduced 
by Zanardi {\it et al.} \cite{Zan00}, and introduce a new one as well, 
based exclusively on the shape of a particular probability (density) 
distribution: that for finding a state with a given entanglement change 
$\Delta E$, measured in terms of the so called entanglement of 
formation \cite{WO98}. We will see that the distribution obtained by 
randomly picking up two states measuring their relative entanglement change is 
optimal in the context of our new measure. Moreover, the two-qubits instance 
will be seen to be rather peculiar in comparison with its counterpart 
for larger dimensions (bipartite systems, 
like two-qudits $N_A \times N_A$, for $N_A$ = 3,4,5 and 6).

Extending the above considerations to mixed states requires the introduction 
of a measure for the simplex of eigenvalues of the matrix $\hat \rho$ instead 
of dealing with pure states distributed according to the invariant 
Haar measure. Rather than mimicking the aforementioned evaluation, which 
could be easily achieved by introducing a proper measure for the generation 
of mixed states, we will generate them in the fashion of 
Refs. (\cite{ZHS98,Z99,ZS01,MJWK01,IH00,BCPP02a,BCPP02b}). In such a 
connection we discuss the action of the exclusive-OR or controlled-NOT 
gate (CNOT in what follows) in the 15-dimensional 
space $\cal S$ of mixed states and compare our results with those 
obtained using the well known Hilbert-Schmidt and Bures metrics \cite{ZyckJPA}.

Also we study numerically how the entanglement is distributed when more than 
two parties are involved (multipartite entanglement). By applying locally 
the CNOT gate to a given pair of two-qubits in a system of pure states 
composed by three or four qubits, we shall study the concomitant distributions 
of entanglement changes among different qubits, pointing out the 
differences between them. Great entanglement changes are appreciated 
as we increase the relevant number of qubits.

The paper is organized as follows: in section II we summarize the optimal 
parametrization of two-qubits gates, which will be useful in order to discuss, 
in section III, the computation of the so called ``entangling power" 
\cite{Zan00} for several quantum gates. In this regard we will introduce a new 
measure for general bipartite states (two-qubits and two-qudits). 
The extension to mixed states is dealt with in section IV by recourse to 
an appropriate heuristic approach that uses the Hilbert-Schmidt and 
Bures metrics. In section V we study some basic properties of the 
distribution of entanglement in multipartite systems and the effects 
produced by two-qubits gates acting upon them. Finally, some conclusions 
are drawn in section VI.



\section{Optimal parametrization of quantum gates for two-qubits systems}

Two-qubits systems are the simplest quantum ones exhibiting 
the entanglement phenomenon. They play a fundamental role in 
quantum information theory. There remain still some features 
of these systems, related to the phenomenon of entanglement, that 
have not yet been characterized in enough detail, as for instance, 
the manner in which $P(\Delta E)$, the probability of generating a 
change $\Delta E$ associated to the action of these operators, 
is distributed under the action of certain quantum gates. In this 
vein it is also of interest to express the general quantum two-qubits 
gate in a way as compact as possible, i.e., to find an optimal parametrization.

Since any quantum logical gate acting on a two-qubits system can be 
expressed in the form \cite{VHC02},

\begin{equation} 
\label{lambdecomp1} 
\left( v_1 \otimes v_2 \right) \, \exp 
\left[-i
\sum_{i=1}^3 \lambda_k \sigma_k \otimes \sigma_k \right] \, 
\left( w_1 \otimes w_2  \right), 
\end{equation} 

\noindent where the transformations $v_{1,2}$ and $w_{1,2}$ act only on one of 
the two qubits, and $\sigma_k$ are the Pauli matrices. Note that it is always 
possible to chose the $\lambda$-parameters in such a way that

\begin{eqnarray}
\lambda_1 \ge \lambda_2 \ge |\lambda_3|, \cr \lambda_1, \lambda_2 \in
[0,\pi/4], \cr \lambda_3 \in (-\pi/4, \pi/4], 
\end{eqnarray}

\noindent and consider the parameterized unitary transformation

\begin{equation} \label{lambdecomp2} \tilde U_{(\lambda_1,\lambda_2,\lambda_3)} 
\, = \, \exp \left[-i \sum_{i=1}^3 \lambda_k \sigma_k \otimes \sigma_k \right]. 
\end{equation} 

From previous work \cite{VHC02,PhysicaApending} we know that the unitary 
transformations (\ref{lambdecomp1}) and (\ref{lambdecomp2}) share the same 
probability distribution $P(\Delta E)$ of entanglement changes. Consequently, 
the $P(\Delta E)$-distribution generated by any quantum logical gate 
acting on a two-qubits system coincides, for appropriate values of the 
$\lambda$-parameters, with the distribution of entanglement changes 
associated with a unitary transformation of the form (\ref{lambdecomp2}). 
This means that the set of all possible $P(\Delta E)$-distributions for 
two-qubits gates constitutes, in principle, a three-parameter family 
of distributions.  

As an example of this equivalence between gates, let us consider 
the CNOT and $\hat U_{\theta}$ gates

\begin{equation} \label{CNOT}
CNOT \, = \,
\left(
\begin{array}{cccc}
1 & 0 & 0 & 0 \\
0 & 1 & 0 & 0 \\
 0 & 0 & 0 & 1 \\
 0 & 0 & 1 & 0
\end{array}
\right), U_{\theta} \, = \, \left(
\begin{array}{cccc}
 1 & 0 & 0 & 0 \\
 0 & 1 & 0 & 0 \\
 0 & 0 & \cos(\theta) & \sin(\theta) \\
 0 & 0 & -\sin(\theta) & \cos(\theta)
\end{array}
\right).
\end{equation} 

\noindent It can be shown \cite{PhysicaApending} that, defining the 
auxiliary gates

\begin{equation}
U_{LA} \, = \, \left(
\begin{array}{cc}
 1 & 0 \\
 0 & e^{i\pi/2}
\end{array}
\right) \otimes
 \left(
\begin{array}{cc}
e^{-i\pi/2} &  0 \\
   0        &  1
 \end{array}
 \right), U_{LB} \, = \, \left(
\begin{array}{cc}
 1 & 0 \\
 0 & 1
\end{array}
\right)
\otimes
 \left(
\begin{array}{cc}
e^{i\pi/2} &  0 \\
   0        &  1
 \end{array}
 \right),
\end{equation} 

\noindent the formal relation 
$U_{\pi/2} \, = \, U_{LA} \, {\rm CNOT} \, U_{LB}$ holds, implying 
that both CNOT and $U_{\pi/2}$ share the same $P(\Delta E)$-distribution.

We have explored the two-qubits space by means of a Monte Carlo simulation 
\cite{ZHS98,Z99,PZK98} and in Fig. 1 we depict the action of several gates acting on 
two-qubits pure states, as described by different values of the 
vector $(\lambda_1,\lambda_2,\lambda_3)$. We see how different 
the associated entanglement probability distributions are. In point of fact, 
the CNOT gate (solid line) is equivalent (on average) to $(\pi/4,0,0)$. 
Curve 1 corresponds to $\lambda$ = ($\pi/4$, $\pi/8$, 0), curve 2 
to ($\pi/4$, $\pi/8$, $\pi/16$), curve 3 to ($\pi/4$, 0, 0), 
curve 4 to ($\pi/4$, $\pi/8$, -$\pi/8$), and curve 5 to 
($\pi/8$, $\pi/8$, $\pi/8$). All these gates have the common property 
that they reach the extremum $|\Delta E|=1$ change if have given 
the appropriate ${\bf \lambda}$ vectors. This is not the case for other gates 
like the $U_{\pi/4}$ one \cite{BPCP03}. The vertical dashed line represents 
any gate that can be mapped to the identity $\hat I$, so that no change in 
the entanglement occurs (we get a delta function $\delta(\Delta E)$).


\section{Quantum gates' entangling power: qubits and qudits}

As stated, a quantum gate (QG), represented by a unitary 
transformation $\hat U$, changes the entanglement of a given state. As 
a matter of fact, we may think of the QG as an ``entangler". This 
particular transformation represents the abstraction of some physical 
interaction taking place between the different degrees of freedom of 
the pertinent system. A natural question then arises: how good 
a quantum gate is as an entangler?, or in other words, can we quantify 
the set of quantum gates in terms of a certain ``entanglement capacity"? 
The question is of some relevance in Quantum Information. A quantum gate 
robust against environmental influence becomes specially suitable 
in the case of networks of quantum gates (quantum circuits, quantum 
computer, etc) as described by Zanardi {\it et al.} \cite{Zan00}, 
where the so called ``entangling power" $\epsilon_P(\hat U)$ of a 
quantum gate $\hat U$ is defined as follows

\begin{equation} \label{eP}
  \epsilon_P(\hat U) \, \equiv \, \overline{E\big((\rho_A\otimes\rho_B) 
  \hat U (\rho_A\otimes\rho_B)^{\dag}\big)}^{\rho_A,\rho_B},
\end{equation}

\noindent where the bar indicates averaging over all (pure) product states 
in a bipartite quantum state described by 
$\rho_{AB}=\rho_A\otimes\rho_B \in \cal H=\cal H_A \otimes \cal H_B$ and 
$E$ represents a certain measure of entanglement, in our case the 
entanglement of formation, that, in the case of pure states becomes just 
the binary von Neumann entropy of either reduced state 
$E(\rho_{AB})=-$Tr$(\rho_A \log_2 \rho_A)=-$Tr$(\rho_B \log_2 \rho_B)$. 
It greatly simplifies the numerics of our study to assume 
that the separable states $\rho_{AB}$ are all equally likely. 
The corresponding (special) form of (\ref{eP}) exhibits the advantage 
that it can be generalized to any dimension for a bipartite system. 
In our case, we are mostly interested in  two-qubits systems 
(the $2 \times 2$ case). In \cite{Zan00} the concept of {\it optimal} gate 
is introduced, where by {\it optimal} one thinks of a gate that makes 
(\ref{eP}) maximal. It is shown there that the CNOT gate is an optimal gate.

Let us suppose now that we make use of the special parameterization 
$\cal{P}$ given in section II for the unitary transformations $U(N)$. 
In the case of the CNOT gate, it was clear that $\cal{P}$ is 
equivalent (on average) to the $(\pi/4,0,0)$ gate. This fact allow us 
to see how the entangling power (\ref{eP}) evolves when we perturb 
the CNOT gate in the form  $(\pi/4,x,x)$, $x$ being a continuous parameter. 
To such an end we numerically generate {\it separable} \cite{comentariPPT} states 
$\rho_A\otimes\rho_B$ according to the Haar measure on the group 
of unitary matrices $U(N)$ that induces a unique and uniform 
measure $\nu$ on the set of pure states of two-qubits ($N=4$) 
\cite{ZHS98,Z99,PZK98}. The corresponding results are shown in Fig. 2. 
Every point has been obtained averaging a sampling of $10^9$ states, 
so that the associated error is of the order of the size of the symbol. 
It is clear from the plot that large deviations imply a smaller 
entangling power $\epsilon_P(CNOT_{pert.})$. Notice that a small 
perturbation around the origin (CNOT gate) {\it increases} 
the entangling power. This fact leads us to conclude that, in the 
space of quantum gates, and in  the vicinity of an optimal gate, there exists 
an infinite number of optimal gates. On the other hand, if we perturb a quantum 
gate which is not optimal, like $(\pi/8,x,x)$, any deviation, no matter how 
small, will lead to an increasing amount of the entangling power $\epsilon_P$. 
This latter case is depicted in the inset of Fig. 2.

It is argued in \cite{Zan00} that the two-qubits case presents 
special statistical features, as far as the entangling power 
is concerned, when compared to $N_A \times N_A$ systems (two-qudits). 
We investigate this point next, not by making use of any 
quantum gate, or by recourse to Eq. (\ref{eP}). What we do instead 
might be regarded a ``no gate action": we look at the probability 
(density) distribution $P_{R}$ obtained by randomly picking up two 
pure states generated according to the Haar measure in $N_A \times N_A$ 
dimensions, and determine then the relative entanglement change 
$\Delta E$ in passing form one of these states to the other. 
The distribution $P_{R}$ is \cite{PhysicaApending}

\begin{equation} \label{randomint}
P_{R}(\Delta E) \, = \, \int^{1-|\Delta E|}_0 \, dE \, P(E) \, P(E+|\Delta E|).
\end{equation}

\noindent The distribution $P_R(\Delta E)$ is thus related to the probability 
density $P(E)$ of finding a quantum state with entanglement $E$. Notice that 
the above expression holds for any states space measure 
invariant under unitary transformations and for any bipartite quantum system
 consisting of two subsystems described by Hilbert spaces of the same 
dimensionality. We must point out that the entanglement is measured for every 
two-qudits in terms of $E=S(\rho_{A})/\log(N_{A})$, 
where $S$ is the von Neumann entropy, so that it ranges from 0 to 1 
($N_A$ is the dimension of subsystem $A$). The resulting distributions 
are depicted in Fig. 3. The five curves represent the 
$2\times2, 3\times3, 4\times4, 5\times5$ and $6\times6$ systems. A first 
glance at the corresponding plot indicates a sudden change in the available 
range of $\Delta E$. The width of our probability distribution is rather large 
for two-qubits and it becomes narrower as we increase the dimensionality 
of the system. With this fact in mind, one may propose the {\it natural width} 
of these distributions as some measure of its entangling power. We choose 
the maximum spread of the distribution in $\Delta E$ at half its maximum 
height $P(0)$. If we use this definition of entangling power $W_{\Delta E}$, 
Fig. 3 provides numerical evidence for the peculiarity of the 
two-qubits instance. One may dare to conjecture, from inspection, 
that for large $N_A$, $W_{\Delta E}$ decays following a 
power law: $W_{\Delta E} \sim 1/N_A^{\alpha}$.



\section{Two-qubits space metrics and the entangling power of a quantum gate}

So far we have considered the QG ``entangling power" as applied to the case 
of pure states of two-qubits. In order to do so, it has been sufficient to 
generate pure states according to the invariant Haar measure. In passing to 
mixed two-qubits states, the situation becomes more involved. 
Mixed states appear naturally when we consider a pure state that is decomposed 
into an statistical mixture of different possible states by environmental 
influence (a common occurrence). It may seem somewhat obvious to extend 
to mixed states the previous study of the entangling power of a certain 
quantum gate by following the steps given by formula (\ref{eP}). Instead, 
we will consider a heuristic approach to the problem.

The space of mixed states $\cal S$ of two-qubits is 15-dimensional, 
which implies that it clearly possesses non-trivial properties, 
a systematic survey of which can be found in 
\cite{ZHS98,Z99,PalmaCorr,PalmaCorr2}. In general, the space ${\cal S}$ 
of all (pure and mixed) states $\hat \rho$ of a quantum 
system described by an $N$-dimensional Hilbert space can be 
regarded as a product space ${\cal S} = {\cal P} \times \Delta$ 
\cite{ZHS98,Z99}. Here $\cal P$ stands for the family of all 
complete sets of orthonormal projectors $\{ \hat P_i\}_{i=1}^N$, 
$\sum_i \hat P_i = I$ ($I$ being the identity matrix).  $\Delta$ 
is the set of all real $N$-uples $\{\lambda_1, \ldots, \lambda_N\}$, 
with $0 \le \lambda_i \le 1$, and $\sum_i \lambda_i = 1$. The 
general state in ${\cal S}$ is of the form 
$\hat \rho =\sum_i \lambda_i P_i$.  The Haar 
measure on the group of unitary matrices $U(N)$ induces a unique, 
uniform measure $\nu$ on the set ${\cal P}$ \cite{ZHS98,Z99,PZK98}. 
On the other hand, since the simplex $\Delta $ is a subset of a 
$(N-1)$-dimensional hyperplane of ${\cal R}^N$, the standard normalized 
Lebesgue measure ${\cal L}_{N-1}$ on ${\cal R}^{N-1}$ provides a natural 
measure for $\Delta$. The aforementioned measures on $\cal P$ and $\Delta$ 
lead to a natural measure $\mu $ on the set $\cal S$ of quantum 
states \cite{ZHS98,Z99},

\begin{equation} \label{memu}
 \mu = \nu \times {\cal L}_{N-1}.
\end{equation}

 Since we consider the set of states of a two-qubits
 system, our system will have $N=4$.
 All present considerations are based on the assumption
 that the uniform distribution of states of a two-qubit system
 is the one determined by the measure (\ref{memu}). Thus, in our
 numerical computations we are going to randomly generate
 states of a two-qubits system according to the measure (\ref{memu}).
 At the same time, we compute distances between states, which can be 
 evaluated by certain measures \cite{ZyckJPA}. The ones that are considered here
 are the Bures distance

\begin{equation} \label{dBures}
d_{Bures}(\hat \rho_1,\hat \rho_2) \, = \, \bigg(2-2\,Tr 
\sqrt{(\sqrt{\hat \rho_2} \hat \rho_1\sqrt{\hat \rho_2)}}\bigg)^{\frac{1}{2}},
\end{equation}

\noindent and the Hilbert-Schmidt distance

\begin{equation} \label{dHS}
d_{HS}(\hat \rho_1,\hat \rho_2) \, = \, \sqrt{|Tr [\hat \rho_1-\hat \rho_2]^2|}.
\end{equation}

\noindent The goal is to generate unentangled 
states $\hat \rho$ (according to (\ref{memu})) of two-qubits and 
to compute by means of measures (\ref{dBures},\ref{dHS}) the 
average distance reached in $\cal S$ by a final state 
$\hat \rho^{\prime}$, once the CNOT gate 
(\ref{CNOT}) is applied. In other words, we quantify the action of the CNOT gate 
acting on the set ${\cal S}^{\prime}$ of completely separable states. The 
several distances between final (after CNOT) and initial states are computed, 
and a probability (density) distribution is then obtained.

The probability distributions for the Bures and Hilbert-Schmidt distances 
are depicted in Fig. 4 and Fig. 5, respectively. However, one has to 
bear in mind that these absolute distances between states do not take into account 
the fact that the set ${\cal S}^{\prime}$ may have (and indeed such is the case) 
a certain non-trivial geometry, which makes the shape of the convex set 
of separable states ${\cal S}^{\prime}$ highly anisotropic \cite{future work}. 
Therefore, in order to clarify the action of the CNOT gate, we separate the 
set ${\cal S}^{\prime}$ into two parts: I) ${\cal S}^{\prime}_{I}$, which is 
the set of unentangled states inside the minimal separable ball around 
$\frac{1}{4} \hat I$ of radius $d_{min}$, as measured with either 
(\ref{dBures}) or (\ref{dHS}), and II) ${\cal S}^{\prime}_{II}$, which is 
nothing but ${\cal S}^{\prime} - {\cal S}^{\prime}_{I}$. In point of fact, 
$d_{min}$ corresponds to the radius of a hypersphere in 15 dimensions 
whose interior points have Tr($\hat \rho^2$)$\le 1/3$ (\cite{BCPP02b}). 
As seen from Figs. 4 or 5, the first case exhibits a well defined range. 
This is due to the fact that any unitary evolution (CNOT in our case) 
does not change Tr($\hat \rho^2$), so that the CNOT gate cannot 
produce entanglement at all or, in other words, cannot ``move" to any 
extent a state $\hat \rho$ out of ${\cal S}^{\prime}_{I}$. On the other hand, 
CNOT may entangle in ${\cal S}^{\prime}_{II}$ and displace the whole 
distribution to the right. Indeed, if we consider for both graphs 
the total set ${\cal S}^{\prime}$, the concomitant distributions look 
rather alike. The crossing point of the three curves in Figs. 4 and 5 
corresponds to the border defined by $d_{min}^{Bures}$ and $d_{min}^{HS}$, 
respectively.

In view of these results, one may call a QG ``strong" if its entangling power, 
in acting on  a separable state, is great. Thus a semi-quantitative 
strength-measure could be the average value of the distance 
$\overline{d}^{{\cal S}^{\prime}}$ over the whole set of separable states. 
However, it should be pointed out that any definition of entangling power 
for mixed states would turn out to be metric-dependent, i.e., it 
depends on the set of eigenvalues $\Delta$ wherefrom $\hat \rho$ is generated.

\section{Entanglement distribution in multiple qubits systems}

So far we considered logical QGs acting on two-qubits systems. We pass now to 
multipartite ones (nothing strange: the environment can be regarded as a third 
party), composed of many subsystems. We thus deal with a network of qubits, 
interacting with each other, and with a given configuration. More specifically, 
one could consider the set $\cal S$ of pure states 
$\hat \rho = |\Psi\rangle \langle \Psi|$ ``living" in a Hilbert space of 
$n$ parties (qubits) ${\cal H} = \otimes_{i=1}^{n}{\cal H}_i$.  

The usual three party, physically-motivated case, is the two-qubits system 
interacting with an environment which, as a first approximation, could be 
treated roughly as a qubit (two-level system). In any case, the issue of 
how the entanglement present in a given system is ${\it distributed}$ 
among its parties is interesting in its own right. Therefore, it should 
be of general interest to study the general case of multipartite 
networks of qubits on the one hand, while discussing, on the other one, 
how the dimensionality (the number of qubits) affects the distribution 
of the bipartite entanglement between pairs when we apply, locally, 
a certain quantum gate.

In what follows we consider the Coffman ${\it et \, al.}-$approach 
of \cite{distrb} and consider firstly the case of three qubits in a pure 
state $\hat \rho_{ABC}$. An important inequality exists that refers to 
how the entanglement between qubits is pairwise distributed. 
The entanglement is measured by the concurrence squared $C^2$. 
Even though we handle pure states, once we have traced over 
the rest of qubits we end up with mixed states of two qubits, so that 
a measure for mixed states is needed. $C^2$ is related to the entanglement 
of formation \cite{WO98}. It ranges from 0 to 1. The concurrence is given 
by $C \, = \, max(0,\lambda_1-\lambda_2-\lambda_3-\lambda_4)$, 
$\lambda_i, \, (i=1, \ldots 4)$ being the square roots, in decreasing 
order, of the eigenvalues of the matrix $\rho \tilde \rho$, with 
$\tilde \rho \, = \, (\sigma_y \otimes \sigma_y) \rho^{*} 
(\sigma_y \otimes \sigma_y)$. The latter expression has to be 
evaluated by recourse to the matrix elements of $\rho$ computed 
with respect to the product basis. Considering the reduced 
density matrices 
$\hat \rho_{A}=Tr_{BC}\,(\hat \rho_{ABC})$, 
$\hat \rho_{AB}=Tr_{C}\,(\hat \rho_{ABC})$ and 
$\hat \rho_{AC}=Tr_{B}\,(\hat \rho_{ABC})$, the following 
elegant relation is derived:

\begin{equation} \label{d}
C^{2}_{AB} \,+\, C^{2}_{AC} \le 4 \, det \hat \rho_{A} \, (\equiv C^{2}_{A(BC)}),
\end{equation}

\noindent where $C^{2}_{A(BC)}$ shall be regarded as the entanglement of qubit 
$A$ with the rest of the system. In fact, we are more concerned in quantifying 
(do not be confused with distances of the previous section) 
$d_W \, \equiv \, C^{2}_{A(BC)} \, - \, C^{2}_{AB} \,-\, C^{2}_{AC}$. From 
inspection, $d_W$ ranges from 0 to 1 and can be regarded as a legitimate 
multipartite entanglement measure, endowed with certain 
properties \cite{distrb}.

In Fig. 6 the probability (density) function $P(d_W)$ is obtained by generating 
a sample of pure states of three qubits according to the invariant Haar measure, 
as we did for $n=2$ in sections II-III. It is interesting to notice the bias 
of the distribution, and the remarkable fact that numerical evaluation 
indicates that $\overline{d_W} \simeq 1/3$.

Now, suppose that we apply the CNOT gate to the pair of qubits $AB$. 
This means that the unitarity acting on the state $\hat \rho_{ABC}$ 
is described by $\hat U^{CNOT}_{AB} \otimes \hat I_{C}$, where 
$\hat I_X$ is the identity acting on qubit $X$. Making then 
a numerical survey of the action of this operator on the evolution 
of the system we show the concomitant, pairwise entanglement-change 
$\Delta E$ as the probability distributions plotted in Fig. 7a 
(as measured by the entanglement of formation $E$). Two types of 
entanglement are present in the system, namely, the one between 
the pair $AB$, where the gate is applied, and the remaining possibilities 
$AC$ and $BC$, symmetric on average. The solid thick line depicts 
the first kind $AB$, while the second type $AC,BC$ exhibits a 
sharper distribution (dashed line). One is to compare this 
distributions to the one obtained by picking up two states at 
random (solid thin line), which resembles the case of Fig. 3. 
Again, the random case exhibits a larger width for the 
distribution. When compared to the two-qubits CNOT case 
(thin dot-dashed line), we may think of the existence of a third 
party as a rough ``thermal bath" that somehow dilutes the 
entanglement available to the pair $AB$, as prescribed by 
the relation (\ref{d}). This is why the CNOT distribution 
for $n=3$ seems ``sharper" than that for $n=2$. As a matter of fact, 
if we continue increasing the number of qubits present in the system, 
we can numerically check that the generalization of (\ref{d}) 
still holds. In such a (new) instance, the action of the 
CNOT gate is equivalent to the evolution governed by 
$\hat U^{CNOT}_{AB} \otimes \hat I_{C} \otimes \hat I_{D}$. As it is shown in 
Fig. 7b, the 
new distribution of the entanglement changes for $n=4$ in the 
$AB$ pair (dashed line, out of scale) and, as expected, is more peaked than 
for the $n=2$ (dot-dashed line) and, $n=3$ (solid line) cases, reinforcing 
our thermodynamical analogy \cite{PhysicaApending}. If we compute 
their entangling power (EP) with $W_{\Delta E}$, our new measure 
defined in section III, we could conjecture that the EP decreases 
exponentially with the number of qubits $n$ 
($W_{\Delta E}^{n=2} \simeq 0.437, W_{\Delta E}^{n=3} \simeq 0.196, 
W_{\Delta E}^{n=4} \simeq 0.002$).

\section{Conclusions}


In the present work we have  focused attention upon the action of quantum gates as applied to 
multipartite quantum systems and presented the results of a systematic 
numerical survey. In particular, we investigated  
aspects of the quantum gate or unitary operation (acting on two-qubits 
states) as conveniently represented by a vector ${\bf \lambda} \equiv 
(\lambda_1,\lambda_2,\lambda_3)$, visualizing the ``entangling power" of unitary quantum evolution from 
two different perspectives. 

\begin{itemize}
\item{The first one refers to pure 
states of a bipartite system. One has here a well defined formula that 
quantifies the ability of a given transformation $\hat U$ to entangle, on average, 
a given state that pertains to the set ${\cal S}^{\prime}$ of unentangled 
pure states. We have seen 
that the collective of all possible quantum gates, 
as defined by the vector ${\bf \lambda}$, posseses the following property:  in the vicinity of an optimal gate there are infinite  quantum gates 
 which are optimal as well. In addition, we 
introduced a measure of the entangling power above referered to: $W_{\Delta E}$, on the basis of the 
probability (density) distribution (associated with a quantum gate)  
of finding a state that experiences a given change $\Delta E$ in 
its entanglement $E$. A power-law decay is conjectured: 
$W_{\Delta E} \sim 1/N_A^{\alpha}$, $N_A$ being the dimension of the 
subsystem ($N = N_A \times N_A$).}

\item{The second instance deals with mixed states and the metrics of the 
15-dimensional space $\cal S$ of mixed states of two-qubits. We introduce an 
heuristic measure based on an  average distance $\overline d$ obtaind from  
the  distribution of distances between states in $\cal S$, as defined by the 
action of a definite quantum gate acting (again) on the set of unentangled 
states ${\cal S}^{\prime}$.} 
\end{itemize}

Finally, we have studied i) some basic properties of the 
distribution of entanglement in multipartite systems (MS) (network of qubits) 
and ii) the effects 
produced by two-qubits gates acting upon MS. The fact that the entanglement 
between pairs becomes diluted by the presence of third or fourth parties becomes 
apparent from the concomitant distribution of entanglement changes. 
Their natural 
width $W_{\Delta E}$ decreases with the number of parties $n$,  
in what seems to be an exponential fashion.


\acknowledgements This work was partially supported by the MCyT grant 
BMF2002-03241-FEDER, by the Government of the Balearic Islands, and by CONICET 
(Argentine Agency).

\newpage

\noindent {\bf FIGURE CAPTIONS}

\vskip 0.5cm

\noindent

Fig.1- $P(\Delta E)$-distributions generated by the two-qubits quantum gates, 
parametrized in an optimal way. Curve 1 corresponds 
to $\lambda$ = ($\pi/4$, $\pi/8$, 0), curve 2 to ($\pi/4$, $\pi/8$, $\pi/16$), 
curve 3 to ($\pi/4$, 0, 0) (or equivalently to the CNOT gate), 
curve 4 to ($\pi/4$, $\pi/8$, -$\pi/8$) and  curve 5 to 
($\pi/8$, $\pi/8$, $\pi/8$). The vertical line represents any gate 
that can be mapped to the identity $\hat I$. All depicted quantities 
are dimensionless.

\vskip 0.5cm

\noindent

Fig.2- Entangling power $\epsilon_{P}$ of the perturbed CNOT gate, expressed 
in the form of ($\pi/4, x, x$). Small perturbations around this optimal 
gate ($x=0$) find gates which are also optimal (greater $\epsilon_{P}$). 
Large deviations diminish the concomitant $\epsilon_{P}$. 
A perturbed non-optimal gate, like ($\pi/8, x, x$) shown in the inset, 
increases its $\epsilon_{P}$. See text for details. All depicted quantities 
are dimensionless.

\vskip 0.5cm

\noindent

Fig.3- $P(\Delta E)$-distributions generated ($\Delta E$ being the change 
in the entanglement of formation) by randomly choosing the initial and final 
pure two-qubits states ($2 \times 2$), and several two-qudits states 
($N_A \times N_A$, for $N_A=3,4,5,6$). The two-qubits instance appears to 
be a peculiar case. All depicted quantities are dimensionless.

\vskip 0.5cm

\noindent

Fig.4- Probability (density) distributions of finding a state of 
two-qubits (pure or mixed) being sent a distance $d_{Bures}$ away 
from the original state $\hat \rho$, after the action of the 
CNOT gate. All initial states belong to the set ${\cal S}^{\prime}$ 
of separable states. Two regions are defined. See text for details. 
All depicted quantities are dimensionless.  

\vskip 0.5cm

\noindent

Fig.5- Same as in Fig. 4, using the Hilbert-Schmidt distance $d_{HS}$ between 
states. Both figures show similar qualitative features. 
See text for details. All depicted quantities are dimensionless.

\vskip 0.5cm

\noindent

Fig.6- Probability (density) distribution of finding a pure state 
of three-qubits with a given value of $d_W$ (\ref{d}), a measure of 
the distribution of the pairwise entanglement in the system. The 
curve is biased to low values of $d_W$, 
and $\overline{d_W} \simeq 1/3$. All depicted quantities are dimensionless. 

\vskip 0.5cm

\noindent

Fig.7a- $P(\Delta E)$-distributions generated by the CNOT 
quantum gate $\hat U^{CNOT}_{AB} \otimes \hat I_{C}$, acting on the pair 
$AB$ of a pure state of three-qubits. The resulting distribution 
(solid thick line) is to be compared with the one of the pairs $AC,BC$, 
equal on average (dashed line), the random case where no gate is applied 
(solid thin line) and the case of solely two-qubits CNOT gate 
$P(\Delta E)$ distribution (thin dot-dashed line). As compared to 
the three-qubit random instance, it possesses a width slightly 
inferior, being much narrower than in the two-qubits case. 
This fact indicates that the entanglement available to the pair 
$AB$ is diluted by the presence of a third party. 
All depicted quantities are dimensionless. 

\vskip 0.5cm

\noindent

Fig.7b- These distributions result from the action of the CNOT gate 
$\hat U^{CNOT}_{AB}$ on two-qubits ($n=2$, dot-dashed line), 
$\hat U^{CNOT}_{AB} \otimes \hat I_{C}$ on three qubits ($n=3$, solid line), 
and $\hat U^{CNOT}_{AB} \otimes \hat I_{C} \otimes \hat I_{D}$ 
on four qubits ($n=4$, dashed line) pure states. The width of these distributions, 
or entangling power $W_{\Delta E}$ (see text), decreases exponentially 
as the number of qubits is increased. All depicted quantities are dimensionless.


\begin{thebibliography}{}
\bibitem{comentariPPT} Or unentangled states. By construction, 
product states are states with no quantum correlation between parties. 
A general necessary criterion for ascertaining when a state (pure or mixed) 
is entangled or not is given by the so called Positive Partial Transpose
criterion (PPT), first derived by Peres \cite{Peres}. Is is proven 
to be sufficient for $2 \times 2$ and $2 \times 3$ systems \cite{Horos}.
\end{thebibliography}
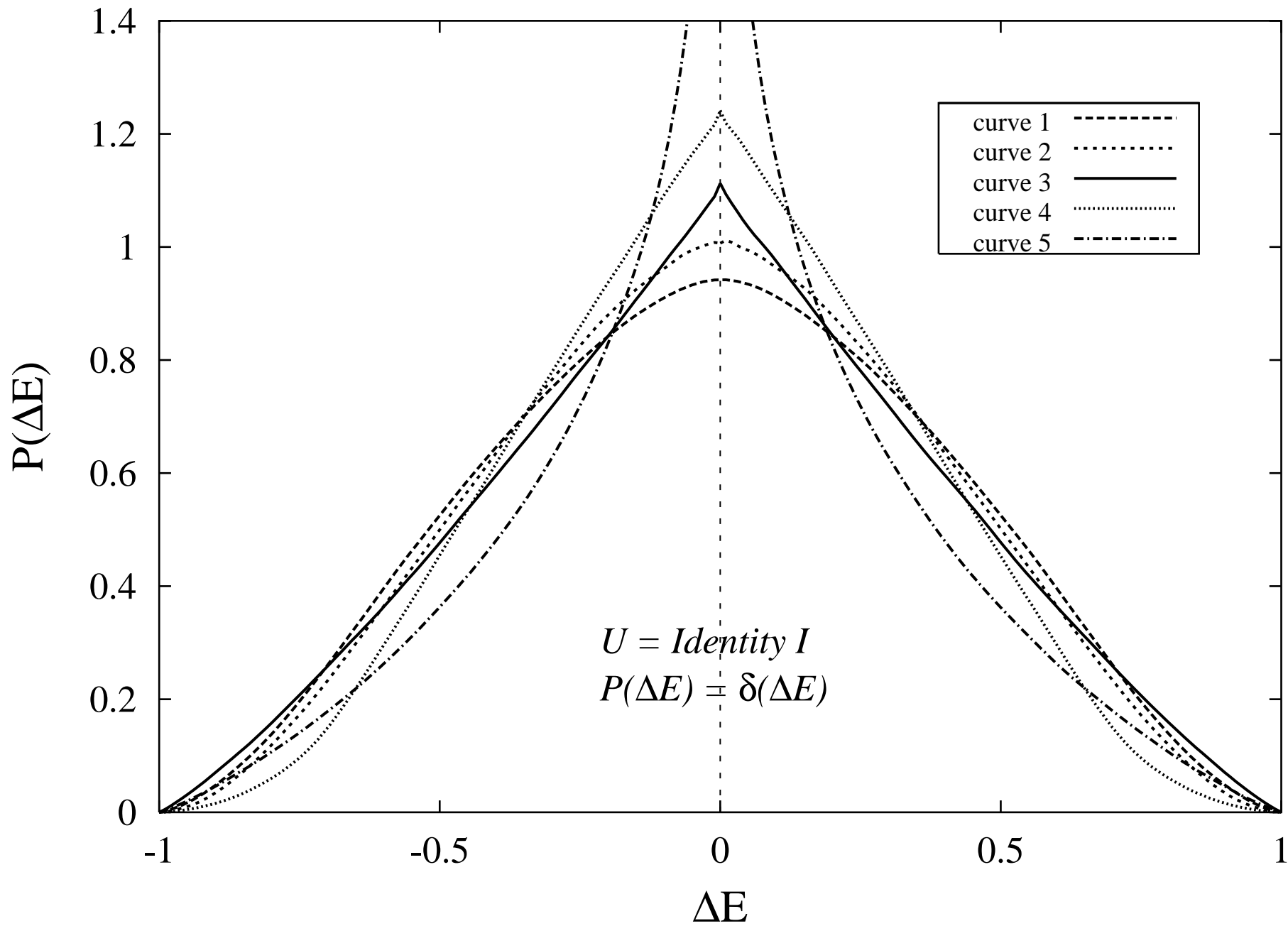
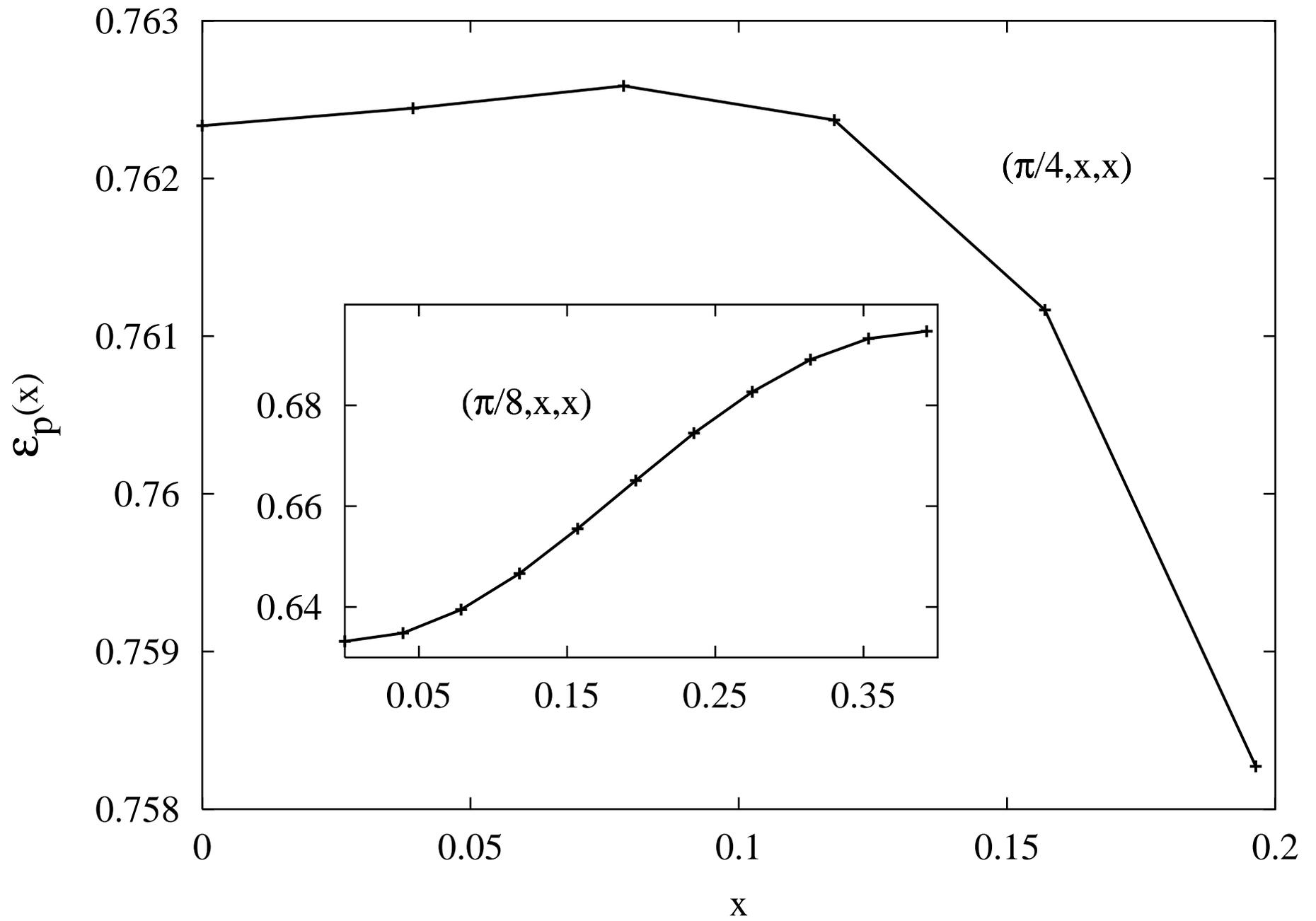
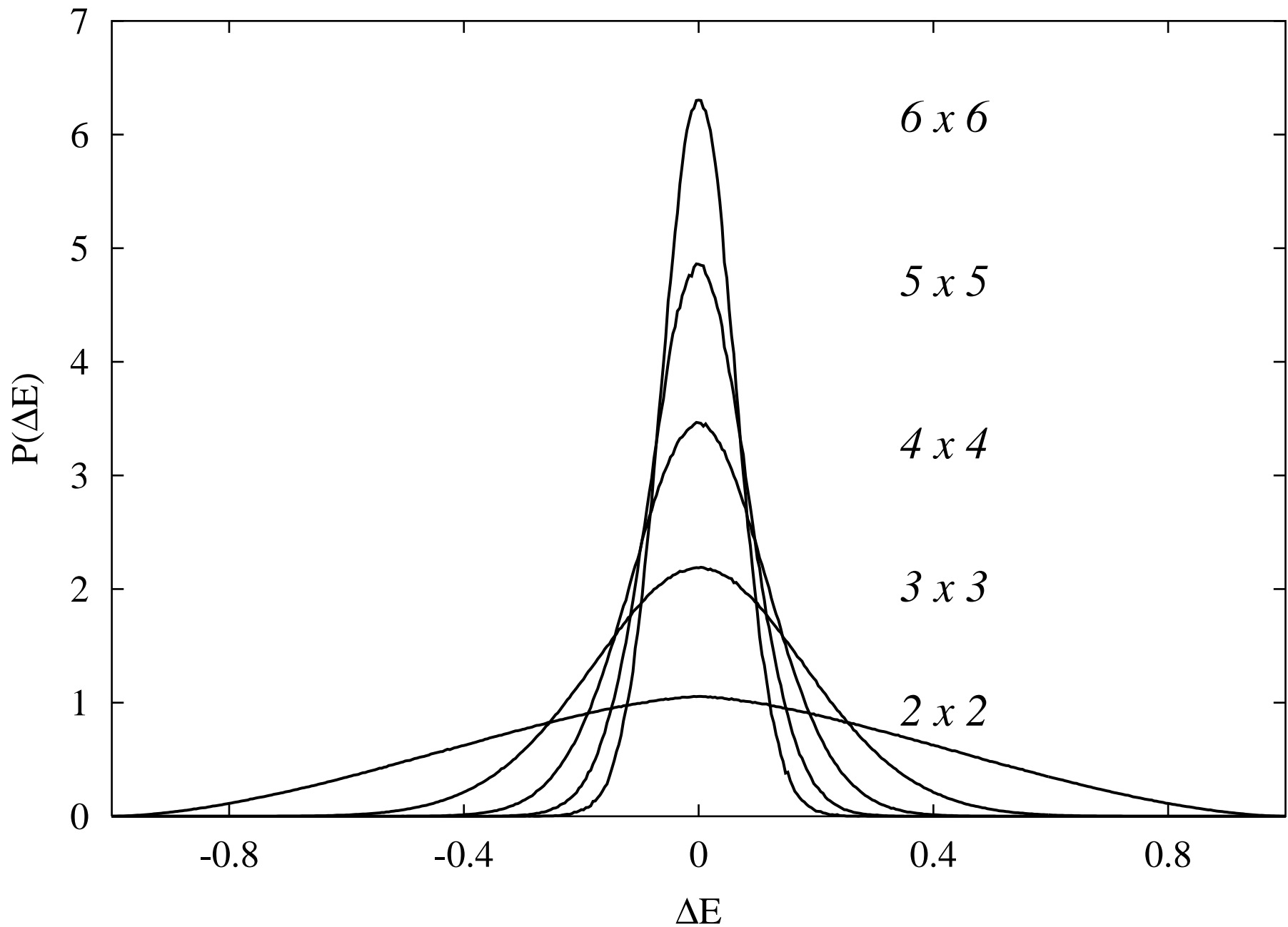
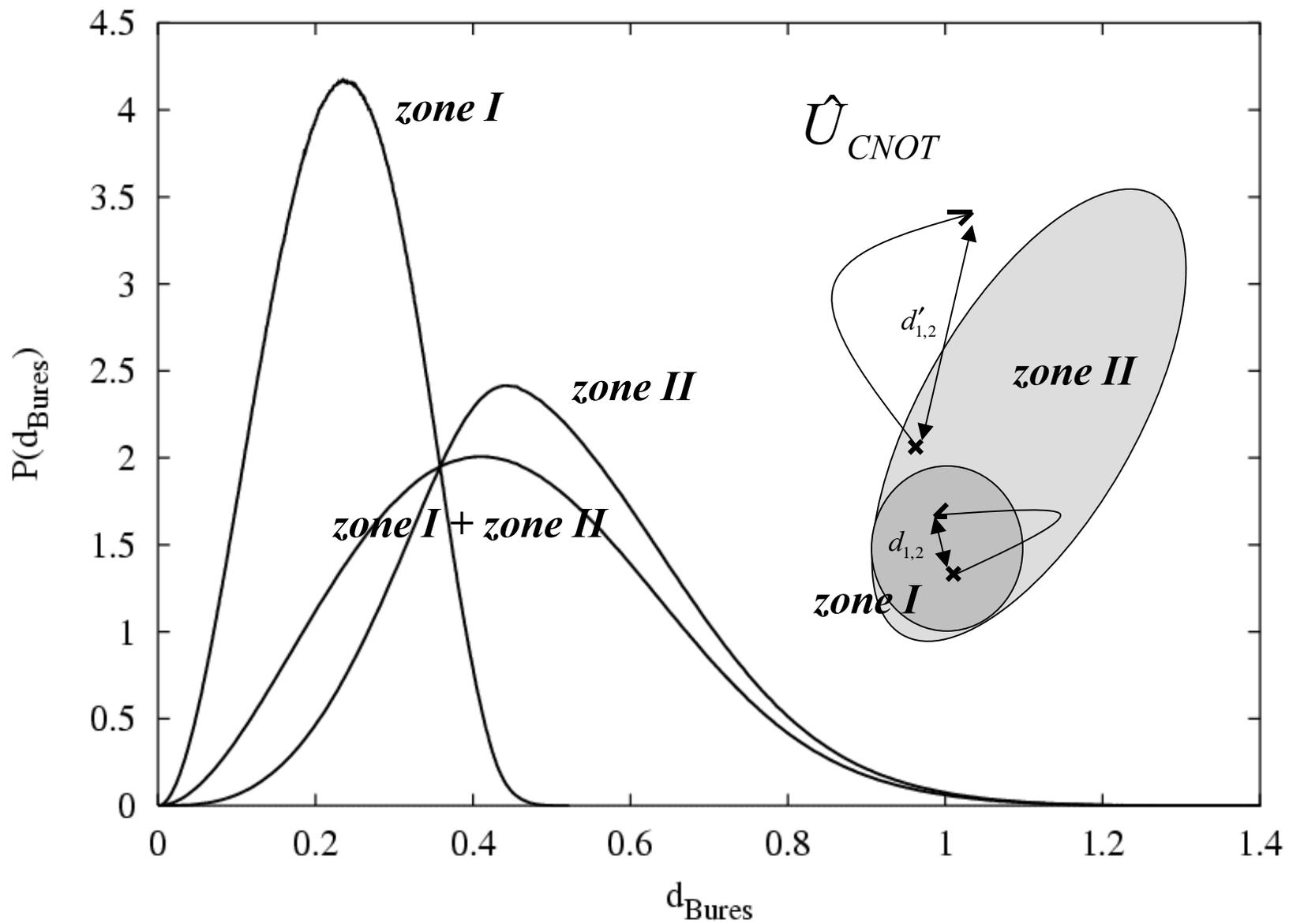
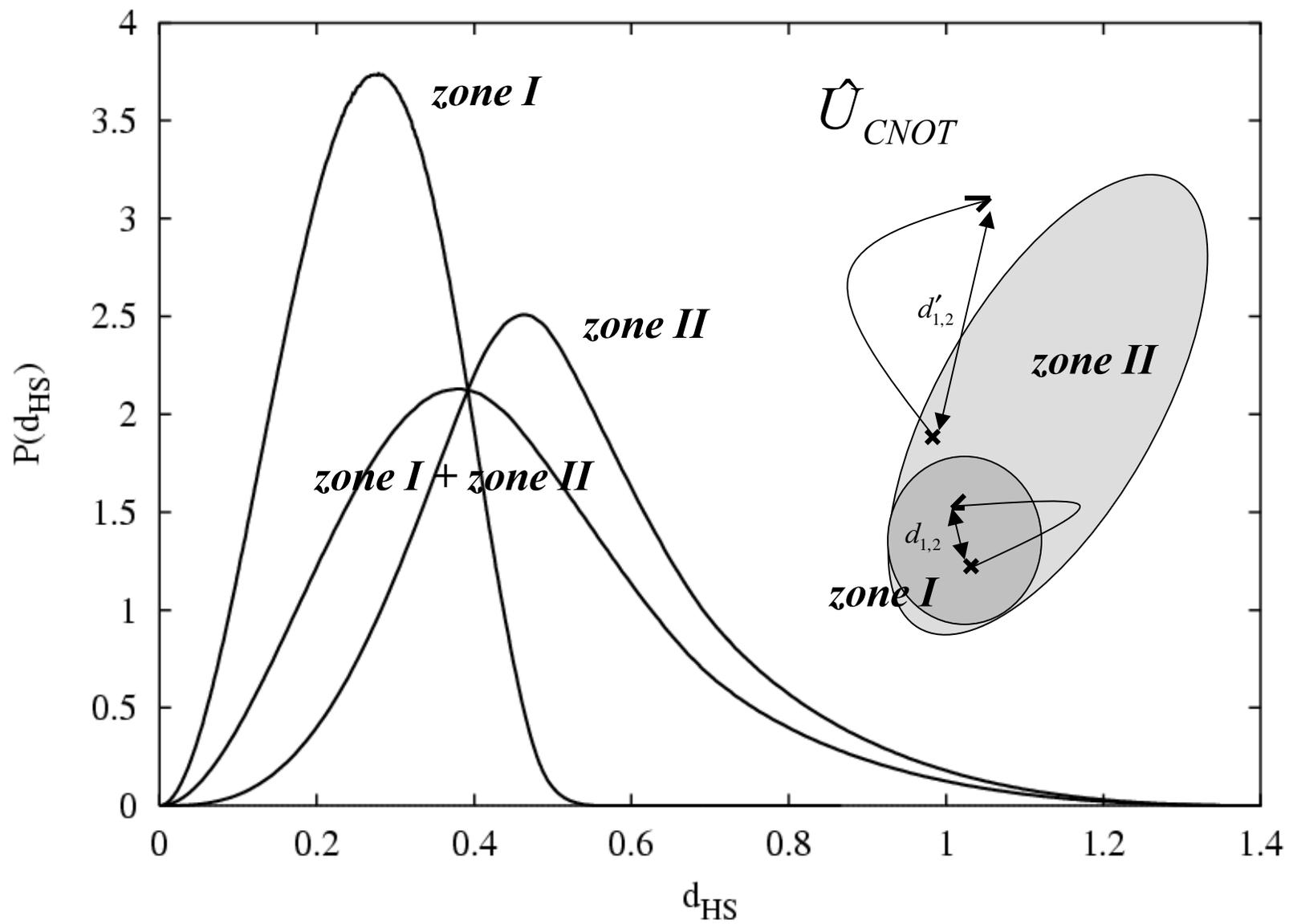
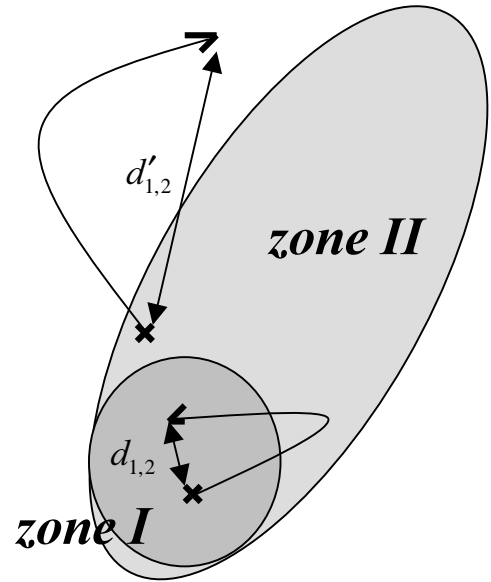
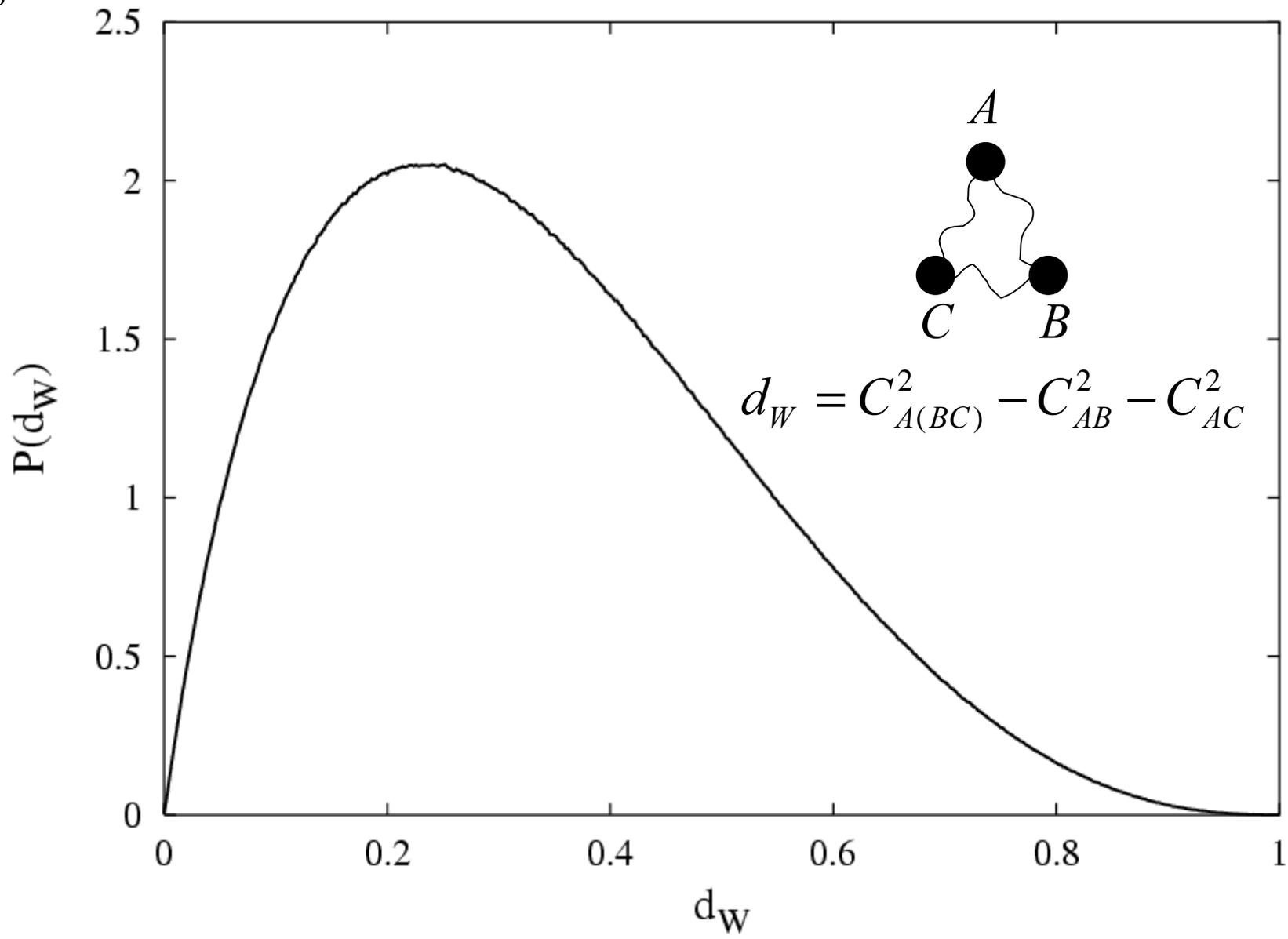
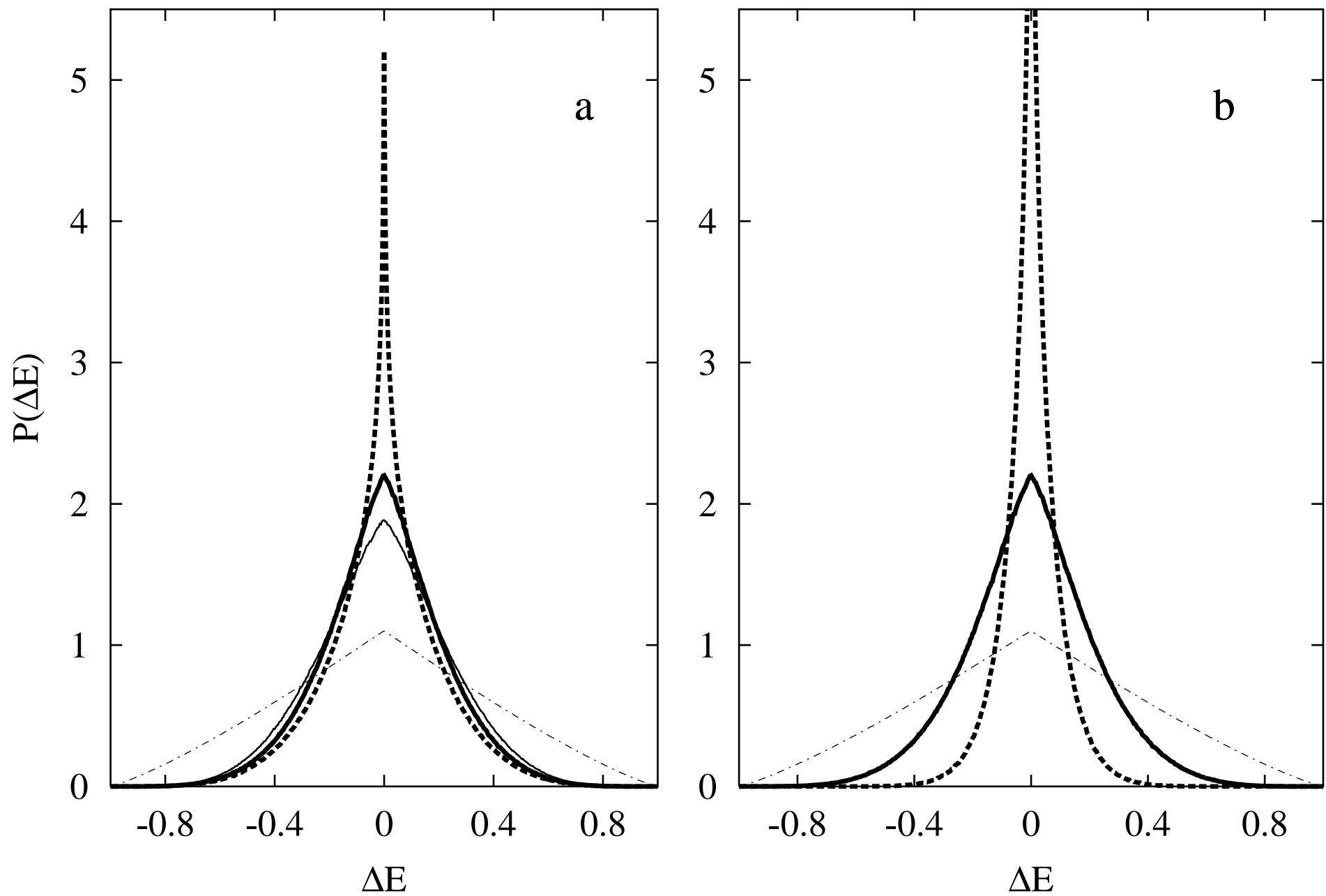
\end{document}